\documentclass[journal,article]{Definitions/mdpi}

\makeatletter
\def\@journal{symmetry} % 或者 def\@journal{none}
\makeatother

\firstpage{1} 
\pubvolume{1}
\issuenum{1}
\articlenumber{0}
\pubyear{2026}
\copyrightyear{2026}
\datereceived{ } 
\daterevised{ } % Comment out if no revised date
\dateaccepted{ } 
\datepublished{ } 
\Title{Continuous-mode Analysis of Improved Two-way CV-QKD}

\Author{Yanhao Sun $^{1}$\orcidA{}, Jiayu Ma $^{1}$\orcidB{}, Xiangyu Wang $^{1,}$*\orcidC{}, Song Yu $^{1}$, Ziyang Chen $^{2,}$*\orcidD{} and Hong Guo $^{2}$\orcidE{}}

\AuthorNames{Yanhao Sun, Jiayu Ma, Xiangyu Wang, Song Yu, Ziyang Chen and Hong Guo}

\address{%
$^{1}$ \quad State Key Laboratory of Information Photonics and Optical Communications, Beijing University of Posts and Telecommunications, Beijing 100876, China\\
$^{2}$ \quad Beijing Key Laboratory of Quantum Sensing and Precision Measurement, School of Electronics, and Center for Quantum Information Technology, Peking University, Beijing 100871, China
}

\corres{Correspondence: xywang@bupt.edu.cn (X.W.); chenziyang@pku.edu.cn (Z.C.)}

\abstract{Continuous-variable quantum key distribution (CV-QKD) enables information-theoretically secure key generation between legitimate parties.
To further enhance system performance, an improved two-way CV-QKD protocol has been proposed, which is accessible in practice and exhibits increased robustness against excess noise.
However, in practical implementations, device nonidealities inevitably drive the optical field from the single-mode regime into the continuous-mode regime.
In this work, we introduce temporal modes to characterize the evolution of optical fields in the improved two-way protocol and establish a security analysis framework for the continuous-mode scenario based on adaptive normalization with calibrated shot-noise unit.
In addition, finite-size effects are taken into account in the analysis.
Our results demonstrate that the improved two-way protocol retains a performance advantage over one-way counterpart.
The analysis provides useful guidance for the practical implementation and performance optimization of improved two-way CV-QKD systems.}


\makeatletter
\let\linenumbers\relax

\makeatother

\begin{document}
\nolinenumbers

%%%%%%%%%%%%%%%%%%%%%%%%%%%%%%%%%%%%%%%%%%

% The order of the section titles is different for some journals. Please refer to the "Instructions for Authors” on the journal homepage.

\section{Introduction}
\label{I}
Quantum key distribution (QKD) \cite{R_QKD1,R_QKD2,R_QKD3,R_QKD4,R_QKD5,R_QKD6} enables two legitimate communicating parties (commonly referred to as Alice and Bob) to establish information-theoretically \cite{R_InformationSecurity} secure keys, effectively addressing the potential threats posed by quantum computing \cite{R_Shor} to classical cryptographic systems. Among various QKD implementations, continuous-variable QKD (CV-QKD) \cite{R_CVQKD1,R_CVQKD2} is naturally compatible with coherent optical communication technologies, making it well suited for large-scale deployment over metropolitan distances. As a result, CV-QKD has attracted significant research interest and has achieved substantial progress in 
theory \cite{R_T1,R_T2,R_T3,R_T4,R_T5,R_T6,R_T7}, 
experimental demonstrations \cite{R_E1,R_E2,R_E3,R_E4,R_E5,R_E6,R_E7,R_E8}, 
network deployment \cite{R_N1,R_N2,R_N3,R_N4}, 
and post-processing techniques \cite{R_P1,R_P2,R_P3,R_P4}.

The maximum transmission distance of CV-QKD is highly sensitive to excess noise \cite{R_MTE}. To enhance the noise tolerance of CV-QKD systems, a two-way protocol has been proposed \cite{R_OTW}. In order to guarantee security, the original two-way protocol requires implementing the tomography of the quantum channels, which is complicated in practice. Subsequently, an improved two-way protocol was introduced \cite{R_ITW}. By considering the purification of the system held by an eavesdropper (Eve), its security can be analyzed using the optimality of Gaussian attacks, which significantly simplifies the security analysis and makes the protocol more feasible for implementation.

However, as CV-QKD increasingly incorporates digital signal processing techniques from coherent optical communication systems \cite{R_DSP1,R_DSP2,R_DSP3}, device nonidealities drive the conventional single-mode optical-field assumption toward a continuous-mode description \cite{R_CM1,R_CM2}. 
Meanwhile, existing security analyses of the improved two-way protocol still rely on the asymptotic scheme \cite{R_AS} that the communicating parties exchange infinitely many signals. To address the practical security of the improved two-way protocol, a new analysis framework is required.

In this work, we introduce temporal modes (TMs) \cite{R_T7,R_TM1,R_TM2,R_TM3} to characterize the evolution of optical fields in the improved two-way protocol and establish a security analysis framework for the continuous-mode scenario based on adaptive normalization with properly calibrated shot-noise unit (SNU). This framework enables a consistent description of continuous-mode interference and detection processes in the improved two-way protocol.
In addition, by invoking the central limit theorem and the maximum likelihood estimation theorem, we analyze the impact of finite-size effects on statistical fluctuations in parameter estimation, leading to a tighter secret key rate under practical conditions.

Numerical simulations for the considered parameter settings show that, the performance of the improved two-way protocol remains superior to that of the one-way protocol. Specifically, the maximum transmission distance is increased by about 24\%, and at a transmission distance of 50 km, the maximum tolerable excess noise is approximately three times higher than that of the one-way protocol.
As the transmission distance increases, the system becomes more sensitive to excess noise, which further highlights the advantage of the improved two-way protocol in long-distance transmission. Overall, our work provides a more practical security and performance analysis of the improved two-way protocol and offers guidance for its implementation and performance optimization.

This paper is organized as follows.
In Section~\ref{II}, we introduce the protocol in the continuous-mode scenario.
In Section~\ref{III}, we analyze the impact of finite-size effects on the secret key rate of the protocol.
In Section~\ref{IV}, numerical simulations are performed and analyzed.
Finally, our conclusions are summarized in Section~\ref{V}.

%%%%%%%%%%%%%%%%%%%%%%%%%%%%%%%%%%%%%%%%%%
\section{Continuous-mode Analysis of Improved Two-way QKD}
\label{II}
In this section, we introduce the improved two-way quantum key distribution protocol in the continuous-mode scenario. 

\subsection{Prepare-and-measure Scheme of the Protocol}
\label{2.1}
\begin{figure}[H]
    \centering
    \includegraphics[width=1.0\linewidth]{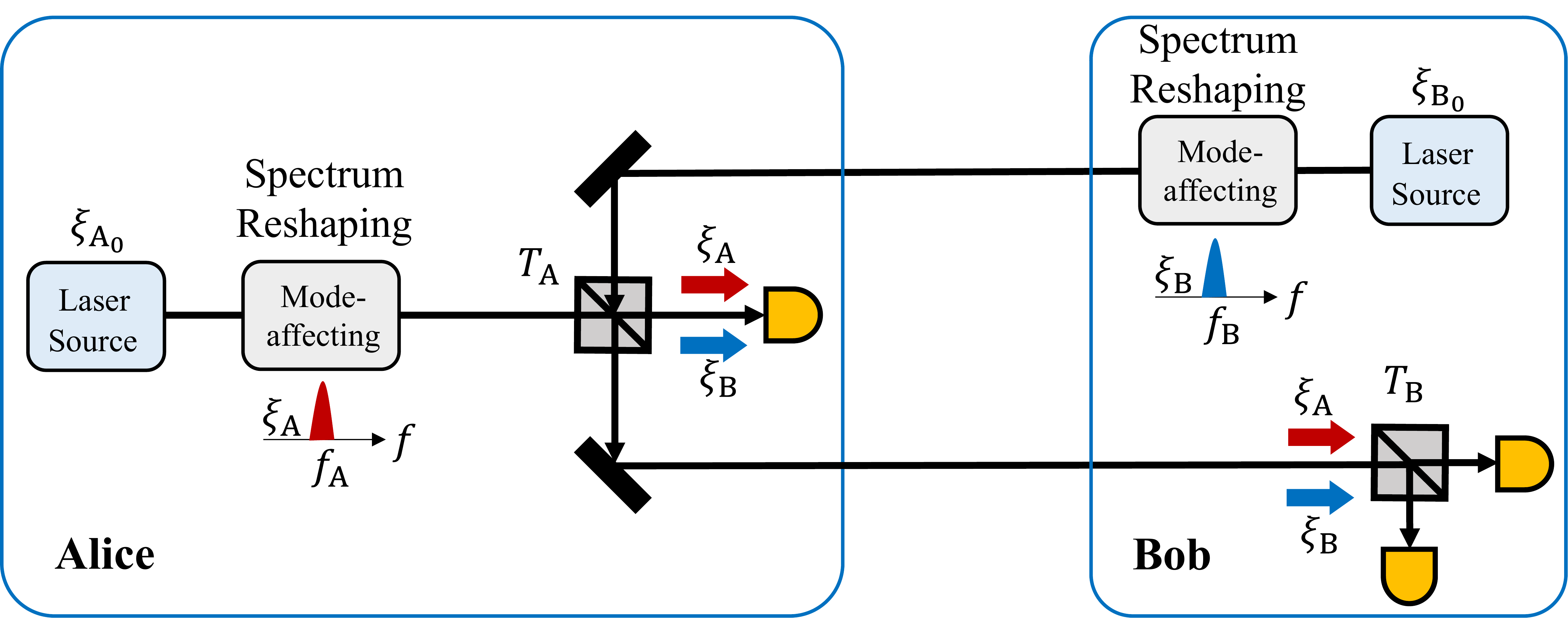}
    \caption{Prepare-and-measure scheme of the improved two-way protocol in the continuous-mode scenario. $\xi$ denotes a wavepacket that contains the temporal and spectral information.}
    \label{FIG1}
\end{figure}

We first present the prepare-and-measure (PM) scheme of the protocol in the continuous-mode scenario, as illustrated in Figure~\ref{FIG1}.

Taking the coherent-state protocol with heterodyne detection as an example, the main steps of the protocol are summarized as follows:

\textbf{1. State preparation.}
In the improved two-way protocol, both legitimate parties prepare coherent states. However, due to the nonidealities of practical laser sources, the states generated by Alice and Bob in realistic systems are no longer single-mode coherent states, but rather continuous-mode coherent states carrying temporal and spectral information. Bob encodes his information onto a continuous-mode coherent state $\left|x_{\mathrm{B}}+i p_{\mathrm{B}}\right\rangle_{\xi_{\mathrm{B}}}$ and sends it to Alice, while Alice also prepares a continuous-mode coherent state $\left|x_{\mathrm{A}}+i p_{\mathrm{A}}\right\rangle_{\xi_{\mathrm{A}}}$. Here, $\xi_{\mathrm{A}}$ and $\xi_{\mathrm{B}}$ denote the envelopes containing the temporal and spectral information of Alice's and Bob's states, respectively.

\textbf{2. State interference.}
After receiving the continuous-mode coherent state $\left|x_{\mathrm{B}}+i p_{\mathrm{B}}\right\rangle_{\xi_{\mathrm{B}}}$ sent by Bob, Alice interferes it with the continuous-mode coherent state $\left|x_{\mathrm{A}}+ip_{\mathrm{A}}\right\rangle_{\xi_{\mathrm{A}}}$ prepared by herself using a beam splitter with transmittance $T_{\mathrm{A}}$.

\textbf{3. State measurement.}
Alice performs homodyne detection on one output of the interference between the two continuous-mode coherent states, while the other output is sent to Bob. Bob performs heterodyne detection on the received state ($T_{\mathrm{B}}=1/2$). Due to the finite bandwidth of practical detectors, Alice and Bob cannot access the full temporal and spectral information of the quantum states. Consequently, both the modulation rate and the format of the modulation signal employed by the transmitter influence the final detection performance.

\textbf{4. Data processing.}
Bob processes his measurement data to estimate Alice's data, after which Bob and Alice perform classical post-processing procedures such as error correction and privacy amplification.

From the above description, it can be observed that if the security analysis of practical systems is still based on the conventional single-mode model, it does not fully capture the mode mismatch between the transmitter and receiver arising from laser nonidealities, modulation imperfections, and detector nonidealities. Moreover, when Alice interferes the continuous-mode coherent state $\left|x_{\mathrm{A}}+ip_{\mathrm{A}}\right\rangle_{\xi_{\mathrm{A}}}$ prepared by herself with the continuous-mode coherent state $\left|x_{\mathrm{B}}+i p_{\mathrm{B}}\right\rangle_{\xi_{\mathrm{B}}}$ prepared by Bob, the interference process inherently involves continuous-mode optical fields, which is also difficult to be described within the single-mode framework. Therefore, to analyze the improved two-way protocol under practical scenarios, it is necessary to introduce new analytical tools.

\subsection{Entanglement-based Scheme of the Protocol}
\label{2.2}
In conventional security analysis, the ideal single-mode coherent state can be represented by the annihilation and creation operators of a single-mode field, $\hat{a}_i$ and $\hat{a}_i^{\dagger}$. However, in practical systems, modulation at the transmitter inevitably introduces additional frequency components, which manifests as a nonuniform temporal waveform in the time domain. In this case, it is necessary to replace the single-mode operators with continuous-mode operators, defined as \cite{R_CM1} $\hat{a}_i \rightarrow \sqrt{\Delta \omega} \hat{a}(\omega)$ and $\hat{a}_i^{\dagger} \rightarrow \sqrt{\Delta \omega} \hat{a}^{\dagger}(\omega)$, where $\Delta \omega$ denotes the mode spacing. Moreover, by applying the Fourier transform, the operator in the time domain can be expressed as $\hat{a}^{\dagger}{ }(t)=({1}/{\sqrt{2 \pi}}) \int d \omega \hat{a}^{\dagger}(\omega) \exp (-i \omega t)$. Combined with the wavepacket $\xi_i(t)$, the photon wavepacket operator can be obtained as \cite{R_OWO}

\begin{equation}
\hat{A}_{\xi_i}^{\dagger}=\int d t \xi_i(t) \hat{a}^{\dagger}(t).
\end{equation}
The annihilation operator $\hat{A}_{\xi_i}$ follows a similar definition. 
Moreover, if $\xi_i(t)$ meets the orthonormalization, then $\hat{A}_{\xi_i}^{\dagger}$ and $\hat{A}_{\xi_i}$ are also known as the temporal mode (TM) field operators. When $\hat{A}_{\xi_i}^{\dagger}$ acts on a vacuum state, it generates a coherent state with an envelope of $\xi_i(t)$.

After describing the continuous-mode states prepared at the transmitter using TM, the receiver of the system that incorporates continuous-mode effects can likewise be characterized within the TM framework. 
As the continuous-mode coherent states sent by the transmitter carry temporal and spectral information, while the nonideal detectors at the receiver are unable to extract all the information from an arbitrary and unknown temporal wavepacket. In addition, the digital signal processing (DSP) techniques employed at the receiver involve sampling and processing of multi-point measurement outcomes. 
To properly describe the final output of the receiver, it is necessary to perform adaptive shot-noise unit (SNU) normalization according to the specific configuration.
By configuring the local oscillator (LO), filters, sampling procedures, and DSP algorithms in practical systems, the SNU normalization factor of the receiver output can be obtained as \cite{R_T7}
\begin{equation}
\sigma_{\mathrm{SNU}}=\sqrt{\frac{\mu_{\mathrm{LO}}}{\Delta t_{\mathrm{s}}^2} \int\left|\xi_{\mathrm{LO}}(\tau)\right|^2\left[G_{\mathrm{DSP}}^{t_j}(\tau)\right]^2 d \tau},
\end{equation}
where $\mu_{\mathrm{LO}}$ denotes the average number of photons contained in an envelope $\xi_{\mathrm{LO}}(t)$ for a pulsed LO, and $\Delta t_s$ denotes the sampling time interval.
$G_{\mathrm{DSP}}(\tau)=\sum_{k=1}^N f_{\mathrm{DSP}}^k g\left(t_k-\tau\right)$, where $g(t)$ denotes the detector's impulse response function, $k$ denotes the sampling points, and $f_{\mathrm{DSP}}^k$ denotes the coefficients determined by the specific DSP algorithm.

After performing adaptive normalization on the receiver output, the normalized photon wavepacket function is defined as
\begin{equation}
\Xi_{\mathrm{DSP}}^{t_j}(\tau)=\frac{\xi_{\mathrm{LO}}(\tau) G_{\mathrm{dsp}}^{t_j}(\tau) \exp \left(-i \omega_{\mathrm{LO}} \tau\right)}{\sqrt{\int d \tau\left|\xi_{\mathrm{LO}}(\tau)\right|^2\left[G_{\mathrm{dsp}}^{t_j}(\tau)\right]^2}},
\end{equation}
we can further define its creation operator as
\begin{equation}
\hat{A}_{\Xi_{\mathrm{DSP}}^{t_j}}^{\dagger}=\int d \tau \bar{\Xi}_{\mathrm{DSP}}^{t_j}(\tau) \hat{a}^{\dagger}(\tau) .
\end{equation}

In the TM framework, the TM associated with the measured quantum state is denoted as $\xi\text{-TM}$, while the effective TM defined by the receiver is denoted as $\Xi_{\mathrm{DSP}}\text{-TM}$. The continuous-mode measurement can be regarded as a mode projection process. A third mode $\Psi_{\perp}\text{-TM}$ can be constructed via Gram–Schmidt orthogonalization,  which is derived from $\Xi_{\mathrm{DSP}}\text{-TM}$ and orthogonal to $\xi\text{-TM}$.
The corresponding decomposition of the creation operator is then given by

\begin{equation}
    \hat{A}_{\Xi_{\mathrm{DSP}}}^{\dagger}=\sqrt{\eta_{\mathrm{m}}} \hat{A}_{\xi}^{\dagger}+\sqrt{1-\eta_{\mathrm{m}}} \hat{A}_{\Psi_{\perp}}^{\dagger},    
\end{equation}
where $\eta_{\text{m}}$ denotes the mode-matching coefficient,
\begin{equation}
    \label{match}
    \eta_{\mathrm{m}}=\left|\int d t \Xi_{\mathrm{DSP}}^*(t) \xi(t)\right|^2.
\end{equation}

\begin{figure}[H]
    \centering
    \includegraphics[width=1.0\linewidth]{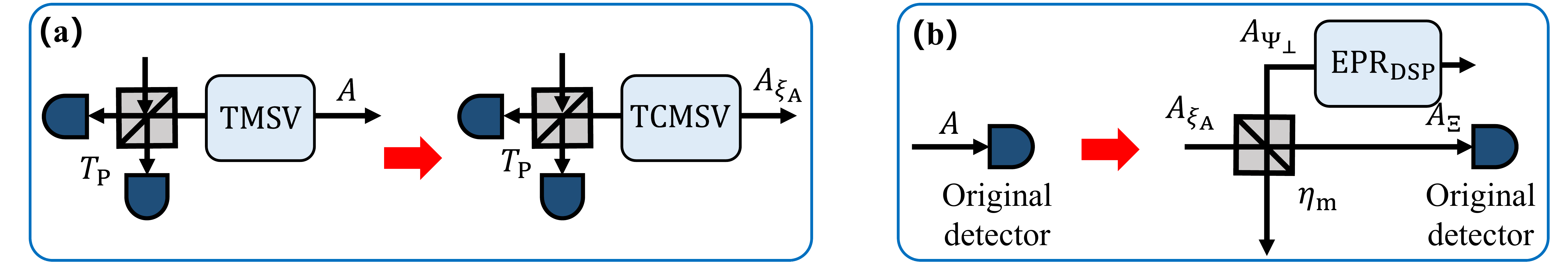}
    \caption{Key equivalence steps between the EB and PM schemes of the protocol. (a) Equivalence at the transmitter. (b) Equivalence at the receiver.}
    \label{FIG2}
\end{figure}

To analyze the security of the PM scheme, we construct an EB scheme that is equivalent to it. The equivalence is established in two aspects.
First, 
\textbf{state-preparation equivalence:} 
performing heterodyne detection on one mode of a two-continuous-mode squeezed vacuum (TCMSV) state is equivalent to preparing a Gaussian-modulated continuous-mode coherent state.
Second, 
\textbf{measurement equivalence:} 
the effect of transmitter–receiver mode mismatch can be equivalently modeled as a reduction in detection efficiency, which can be represented by a beam splitter with transmittance $\eta_{\mathrm{m}}$.
The key steps establishing the equivalence between the EB and PM schemes of the protocol are illustrated in Figure~\ref{FIG2}.

The EB scheme of the protocol is described as follows: 

\textbf{1. State preparation.} 
Bob prepares a TCMSV with variance $V_{\mathrm{B}}$, keeps one mode $B_{1 \xi_{\mathrm{B}}}$, and sends the other mode $B_{\text {out } \xi_{\mathrm{B}}}$ to Alice through the quantum channel. Alice prepares a TCMSV with variance $V_{\mathrm{A}}$, performs heterodyne detection on one of the modes $A_{1 \xi_{\mathbf{A}}}$, and keeps the other mode $A^{\prime}{ }_{\xi_{\mathrm{A}}}$.

\textbf{2. State interference.}
The mode $B_{\text {out } \xi_{\mathrm{B}}}$ sent by Bob evolves through the quantum channel into $A_{\mathrm{in} \xi_{\mathrm{B}}}$. Alice interferes $A_{\mathrm{in} \xi_{\mathrm{B}}}$ with the previously retained mode $A^{\prime}{ }_{\xi_{\mathrm{A}}}$ using a beam splitter with transmittance $T_{\mathrm{A}}$, resulting in the output modes $A_{2 \xi}$ and $A_{\text {out} \xi}$. Alice keeps the mode $A_{2 \xi}$ and sends the mode $A_{\text{out}\xi}$ back to Bob. 
The mode $A_{\text{out}\xi}$ sent by Alice evolves through the channel into $B_{2 \xi}$.

\textbf{3. State measurement.}
Alice performs homodyne detection on the mode $A_{2 \xi}$. At this stage, due to the interference of continuous-mode optical fields, it is necessary to separately consider the mode-matching coefficients between the detection mode at Alice and the modes originating from Bob and Alice, denoted by $\eta_{\mathrm{m}}^{\mathrm{BA}}$ and $\eta_{\mathrm{m}}^{\mathrm{AA}}$, respectively.
 Bob performs heterodyne detection on both the mode $B_{2 \xi}$ and the mode $B_{1 \xi_{\mathrm{B}}}$ originally retained by himself. Similarly, owing to the interference of continuous-mode optical fields, the mode-matching coefficients between Bob's detection mode and the modes originating from Bob and Alice, denoted by $\eta_{\mathrm{m}}^{\mathrm{BB}}$ and $\eta_{\mathrm{m}}^{\mathrm{AB}}$, respectively, must be taken into account.

\textbf{4. Data processing.}
Bob processes his measurement data to obtain
$x_{\mathrm{B}}=x_{\mathrm{B}_{2 \xi}}-k x_{\mathrm{B}_{1 \xi_{\mathrm{B}}}}$ and
$p_{\mathrm{B}}=p_{\mathrm{B}_{2 \xi}}+k p_{\mathrm{B}_{1 \xi_{\mathrm{B}}}}$,
where $k=T \sqrt{T_{\mathrm{A}}} \sqrt{\eta_{\mathrm{m}}^{\mathrm{BB}}} \sqrt{V_{\mathrm{B}}-1} / \sqrt{V_{\mathrm{B}}+1}$. Bob then uses his measurement results to estimate Alice's values. After completing data reconciliation, Bob and Alice perform classical post-processing procedures such as error correction and privacy amplification.

From the above description of the EB scheme, it can be seen that, compared with the single-mode case, the continuous-mode model requires careful consideration of the mode-matching process between the transmitter and receiver. In particular, owing to the interference of continuous-mode optical fields, four mode-matching coefficients between the transmitter and receiver must be taken into account simultaneously. In Section~\ref{3.2}, we will present the final covariance matrix between Alice and Bob by incorporating the effects of finite-size statistics.

%%%%%%%%%%%%%%%%%%%%%%%%%%%%%%%%%%%%%%%%%%
\section{Finite-size Effects in the Improved Two-way Protocol}
\label{III}
In practical implementations of the protocol, the two legitimate parties can exchange only a finite amount of data for the post-processing procedure, which leads to increased statistical fluctuations in sampling-based estimations and prevents them from accurately estimating the channel parameters. In this section, we analyze the impact of finite-size effects on parameter estimation and present the secret key rate formula of the improved two-way protocol when finite-size effects are taken into account.

\subsection{Theoretical Analysis of Finite-Size Effects}
\label{3.1}
\subsubsection{Finite-size Statistical Fluctuations}
\label{3.1.1}
When finite-size effects are taken into account, Alice and Bob cannot precisely characterize the properties of the quantum channel. They need to select $m$ signals from the exchanged $N$ signals for parameter estimation, leaving $n = N - m$ signals for key generation.
In this case, the secret key rate of the protocol should be modified as \cite{R_FSK}
\begin{equation}
K_{\mathrm{finite}}=\frac{n}{N}\left[\beta_{\mathrm{R}} I_{\mathrm{AB}}-\left(S_{\mathrm{BE}}\right)_{\epsilon_{\mathrm{PE}}}-\Delta(n)\right],
\label{keyrate}
\end{equation}
where $I_{\mathrm{AB}}$ denotes the mutual information between Alice and Bob, and $S_{\mathrm{BE}}$ denotes the information leaked from Bob to Eve.
The parameter $\epsilon_{\mathrm{PE}}$ denotes the probability of failure in parameter estimation, and the true channel parameters lie within a certain confidence interval around the estimated values with probability $1 - \epsilon_{\mathrm{PE}}$.
The parameter $\Delta(n)$ is related to the security of privacy amplification, and its value is given by
\begin{equation}
\Delta(n) \equiv\left(2 \operatorname{dim} H_X+3\right) \sqrt{\frac{\log _2(2 / \bar{\epsilon})}{n}}+\frac{2}{n} \log _2\left(1 / \epsilon_{P_A}\right),
\end{equation}
where $H_X$ represents the Hilbert space dimension of the variable $x$ in the raw key, 
in the CV protocol, the $\operatorname{dim} H_X$ is set to 2. 
The smoothing parameter $\bar{\epsilon}$ and the privacy amplification parameter $\epsilon_{\mathrm{PA}}$ are intermediate variables, with their optimal values set to $\bar{\epsilon} = \epsilon_{\mathrm{PA}} = 10^{-10}$.

For a general linear channel, the relationship between the data held by Alice and Bob can be expressed as
\begin{equation}
y=t x+z,
\end{equation}
where $t$ denotes the channel transmittance, with $t = \sqrt{T}$. The variable $z$ represents the equivalent noise and follows a normal distribution with unknown variance of the form $\sigma^2 = 1 + T \varepsilon$. The relationships between $t$ and $S_{\mathrm{BE}}$, as well as between $\sigma^2$ and $S_{\mathrm{BE}}$, are given by

\begin{equation}
\left.\frac{\partial S_{\mathrm{BE}}}{\partial t}\right|_{\sigma^2}<0, \quad 
\left.\frac{\partial S_{\mathrm{BE}}}{\partial \sigma^2}\right|_t>0 .
\end{equation}

Due to finite-size effects, the accuracy with which the legitimate parties can assess Eve's eavesdropping behavior in the channel is reduced. To achieve a high level of security, a worst-case estimation of the eavesdropping must therefore be adopted. Specifically, Alice and Bob infer the minimum value of $t$, denoted as $t_{\min}$, and the maximum value of $\sigma^2$, denoted as $\sigma_{\max}^2$, from the sample data.

By invoking the law of large numbers, the estimators $\hat{t}$ and $\hat{\sigma}^2$ can be approximated as following the distributions: 

\begin{equation}
\hat{t} \sim \mathcal{N}\left(t, \frac{\sigma^2}{\sum_{i=1}^m x_i^2}\right), \quad 
\frac{m \hat{\sigma}^2}{\sigma^2} \sim \chi^2(m-1),
\end{equation}
where $\mathcal{N}$ denotes the normal distribution and $\chi^2$ denotes the chi-square distribution.
Let $t_{\text{real}}$ and $\sigma_{\text{real}}^2$ denote the real (unknown but fixed) values of the channel parameters.
Let $\epsilon_{\mathrm{PE}}$ be the total failure probability of parameter estimation, which is equally allocated to the construction of the one-sided confidence bounds for $t$ and $\sigma^2$, i.e., each one-sided failure probability is taken as $\left[(\epsilon_{\mathrm{PE}})/2\right]$.
As a result, $\mathbb{P}(t_{\text{real}} < t_{\min}) \le \left[(\epsilon_{\mathrm{PE}})/2\right]$
and
$\mathbb{P}(\sigma_{\text{real}}^2 > \sigma_{\max}^2) \le \left[(\epsilon_{\mathrm{PE}})/2\right]$.
By applying the union bound, we obtain:

\begin{equation}
\mathbb{P}\left(t_{\text {real }} \geq t_{\min }, \sigma_{\text {real }}^2 \leq \sigma_{\max }^2\right) \geq 1-\epsilon_{\mathrm{PE}}.
\end{equation}

We then obtain the one-sided lower confidence bound $t_{\min}$ for $t$ and the one-sided upper confidence bound $\sigma_{\max}^2$ for $\sigma^2$:

\begin{equation}
\label{tmin}
t_{\min } \approx \hat{t}-z_{\left[\left(\epsilon_{\mathrm{PE}}\right) / 2\right]} \sqrt{\frac{\hat{\sigma}^2}{m V_{\bmod }}},
\end{equation}

\begin{equation}
\label{sigmamax}
\sigma_{\max }^2 \approx \hat{\sigma}^2+z_{\left[\left(\epsilon_{\mathrm{PE}}\right) / 2\right]} \frac{\hat{\sigma}^2 \sqrt{2}}{\sqrt{m}},
\end{equation}
where $z_{\left[\left(\epsilon_{\mathrm{PE}}\right) / 2\right]}$ satisfies
\begin{equation}
\frac{1}{2}\left\{1-\operatorname{erf}\left(\frac{z_{\left[\left(\epsilon_{\mathrm{PE})} / 2\right]\right.}}{\sqrt{2}}\right)\right\}=\frac{\epsilon_{\mathrm{PE}}}{2},
\end{equation}
where $\mathrm{erf}$ denotes the error function, which is defined as

\begin{equation}
\operatorname{erf}(x)=\frac{2}{\sqrt{\pi}} \int_0^x e^{-t^2} \mathrm{~d} t.
\end{equation}

By replacing the maximum likelihood estimators with their expectation values, $t_{\min}$ and $\sigma_{\max}^2$ are obtained with probability $\left(1 - \epsilon_{\mathrm{PE}}\right)$, yielding the covariance matrix corresponding to the worst-case secret key rate.

\subsubsection{Covariance Matrix in the Ideal Scenario}
\label{3.1.2}
To evaluate the secret key rate under finite-size effects using Equation~\ref{keyrate}, it is necessary to construct the covariance matrix between Alice and Bob for the calculation of $I_{\mathrm{AB}}$ and $\left(S_{\mathrm{BE}}\right)_{e_{\mathrm{PE}}}$.
As described in the EB scheme in Section~\ref{2.2}, after the transmission and measurement of the quantum states, Alice and Bob construct the covariance matrix $\gamma_{B_2 B_1 A_2 A_1}$ from the four modes retained in the protocol. In the ideal scenario, the two parties can exchange an infinite amount of data for parameter estimation, and perfect mode matching is achieved between them. Under these conditions, the corresponding covariance matrix can be expressed as

\begin{equation}
\label{idealM}
\gamma_{B_2 B_1 A_2 A_1}=\left(\begin{array}{cccc}
V_{B_2} \mathrm{I} & C_{B_2 B_1} \sigma_Z & C_{B_2 A_2} \mathrm{I} & C_{B_2 A_1} \sigma_Z \\
C_{B_2 B_1} \sigma_Z & V_{B_1} \mathrm{I} & C_{B_1 A_2} \sigma_Z & 0 \\
C_{B_2 A_2} \mathrm{I} & C_{B_1 A_2} \sigma_Z & V_{A_2} \mathrm{I} & C_{A_2 A 1} \sigma_Z \\
C_{B_2 A_1} \sigma_Z & 0 & C_{A_2 A_1} \sigma_Z & V_{A_1} \mathrm{I}
\end{array}\right),
\end{equation}
where the individual elements are given by

\begin{equation}
\left\{\begin{array}{l}
V_{B_1}=V_{\mathrm{B}} \\
V_{B_2}=T\left(\left(1-T_{\mathrm{A}}\right) V_{\mathrm{A}}+\chi+T T_{\mathrm{A}}\left(V_{\mathrm{B}}+\chi\right)\right) \\
V_{A_2}=T_{\mathrm{A}} V_{\mathrm{A}}+T\left(1-T_{\mathrm{A}}\right)\left(V_{\mathrm{B}}+\chi\right) \\
V_{A_1}=V_{\mathrm{A}} \\
C_{B_2 B_1}=T \sqrt{T_{\mathrm{A}}\left(V_{\mathrm{B}}^2-1\right)} \\
C_{B_1 A_2}=-\sqrt{T\left(1-T_{\mathrm{A}}\right)\left(V_{\mathrm{B}}^2-1\right)} \\
C_{B_2 A_2}=\sqrt{T\left(1-T_{\mathrm{A}}\right) T_{\mathrm{A}}}\left(V_{\mathrm{A}}-T\left(V_{\mathrm{B}}+\chi\right)\right) \\
C_{B_2 A_1}=\sqrt{T\left(1-T_{\mathrm{A}}\right)\left(V_{\mathrm{A}}^2-1\right)} \\
C_{A_2 A_1}=\sqrt{T_{\mathrm{A}}\left(V_{\mathrm{A}}^2-1\right)}
\end{array}\right.,
\end{equation}

\begin{equation}
\chi=\frac{1-T}{T}+\varepsilon.
\end{equation}

In the ideal scenario, the secret key rate can be evaluated using the matrix in Equation~\ref{idealM} together with subsequent matrix transformations. Based on the analyses in Section~\ref{2.2} and Section~\ref{3.1.1}, we will subsequently present the method for calculating the secret key rate in practical systems when both continuous-mode effects and finite-size effects are taken into account.

\subsubsection{Finite-size Adaptive Normalization}
In Section~\ref{3.1.2}, the covariance matrix of Alice and Bob in the ideal scenario has been presented. In practical systems, after adaptive normalization, the transmitted and received signals can be described in terms of TMs, and the mode-matching coefficients between the transmitter and receiver must be taken into account, as listed in Table~\ref{table1}.

\begin{table}[H]
\caption{The mode-matching coefficients between the transmitter and receiver in the improved two-way protocol.\label{table1}}
\centering
\begin{tabular}{cc}
\toprule
\textbf{Coefficient} & \textbf{Description} \\
\midrule
$\eta_{\mathrm{m}}^{\mathrm{AA}}$ & Alice's $\xi_{\mathrm{A}}$-TM vs. Alice's detector's $\Xi_{\mathrm{A}}$-TM \\
$\eta_{\mathrm{m}}^{\mathrm{AB}}$ & Alice's $\xi_{\mathrm{A}}$-TM vs. Bob's detector's $\Xi_{\mathrm{B}}$-TM \\
$\eta_{\mathrm{m}}^{\mathrm{BA}}$ & Bob's $\xi_{\mathrm{B}}$-TM vs. Alice's detector's $\Xi_{\mathrm{A}}$-TM \\
$\eta_{\mathrm{m}}^{\mathrm{BB}}$ & Bob's $\xi_{\mathrm{B}}$-TM vs. Bob's detector's $\Xi_{\mathrm{B}}$-TM \\
\bottomrule
\end{tabular}
\end{table}

At the same time, when finite-size effects are taken into account, worst-case estimations of $T$ and $\epsilon$ in the covariance matrix of Alice and Bob are required. Combined with the theoretical analysis in Sec.~\ref{3.1.1}, after introducing the four mode-matching coefficients into the matrix in Eq.~\ref{idealM}, the covariance matrix corrected by finite-size adaptive normalization (FAN) is given by
\begin{equation}
\left(\gamma_{B_2 B_1 A_2 A_1}\right)_{\mathrm{FAN}}=\left(\begin{array}{cccc}
V_{B_2} \mathrm{I} & C_{B_2 B_1} \sigma_Z & C_{B_2 A_2} \mathrm{I} & C_{B_2 A_1} \sigma_Z \\
C_{B_2 B_1} \sigma_Z & V_{B_1} \mathrm{I} & C_{B_1 A_2} \sigma_Z & 0 \\
C_{B_2 A_2} \mathrm{I} & C_{B_1 A_2} \sigma_Z & V_{A_2} \mathrm{I} & C_{A_2 A 1} \sigma_Z \\
C_{B_2 A_1} \sigma_Z & 0 & C_{A_2 A_1} \sigma_Z & V_{A_1} \mathrm{I}
\end{array}\right)_{\mathrm{FAN}},
\end{equation}
where the individual elements are given by

\begin{equation}
\mathrm{FAN}\left\{\begin{aligned}
& V_{B_1}= V_{\mathrm{B}} \\
& V_{B_2}= T_{\min }\left(1-T_{\mathrm{A}}\right)\left[\eta_{\mathrm{m}}^{\mathrm{AB}} V_A+\left(1-\eta_{\mathrm{m}}^{\mathrm{AB}}\right)\right]+\left(\sigma_{\max }^2-T_{\min }\right)+ \\
& T_{\min } T_{\mathrm{A}}\left[T_{\min } \eta_{\mathrm{m}}^{\mathrm{BB}} V_{\mathrm{B}}+T_{\min }\left(1-\eta_{\mathrm{m}}^{\mathrm{BB}}\right)+\left(\sigma_{\max }^2-T_{\min }\right)\right] \\
& V_{A_2}= T_{\mathrm{A}}\left(\eta_{\mathrm{m}}^{\mathrm{AA}} V_{\mathrm{A}}+1-\eta_{\mathrm{m}}^{\mathrm{AA}}\right)+ \\
&\left(1-T_{\mathrm{A}}\right)\left[T_{\min } \eta_{\mathrm{m}}^{\mathrm{BA}} V_{\mathrm{B}}+T_{\min }\left(1-\eta_{\mathrm{m}}^{\mathrm{BA}}\right)+\left(\sigma_{\max }^2-T_{\min }\right)\right] \\
& V_{A_1}= V_{\mathrm{A}} \\
& C_{B_2 B_1}= T_{\min } \sqrt{T_{\mathrm{A}} \eta_{\mathrm{m}}^{\mathrm{BB}}\left(V_{\mathrm{B}}^2-1\right)} \\
& C_{B_1 A_2}=-\sqrt{T_{\min }\left(1-T_{\mathrm{A}}\right) \eta_{\mathrm{m}}^{\mathrm{BA}}\left(V_{\mathrm{B}}^2-1\right)} \\
& \begin{aligned}
& C_{B_2 A_2}=\sqrt{T_{\min }\left(1-T_{\mathrm{A}}\right) T_{\mathrm{A}}} \\
& {\left[\begin{array}{c}
\sqrt{\eta_{\mathrm{m}}^{\mathrm{AB}} \eta_{\mathrm{m}}^{\mathrm{AA}}} V_{\mathrm{A}}+\sqrt{1-\eta_{\mathrm{m}}^{\mathrm{AB}}} \sqrt{1-\eta_{\mathrm{m}}^{\mathrm{AA}}}- \\
\left(T_{\min } \sqrt{\eta_{\mathrm{m}}^{\mathrm{BB}} \eta_{\mathrm{m}}^{\mathrm{BA}}} V_{\mathrm{B}}+T_{\min } \sqrt{1-\eta_{\mathrm{m}}^{\mathrm{BB}}} \sqrt{1-\eta_{\mathrm{m}}^{\mathrm{BA}}}+\sigma_{\max }^2-T_{\min }\right)
\end{array}\right]}
\end{aligned} \\
& C_{B_2 A_1}= \sqrt{T_{\min }\left(1-T_{\mathrm{A}}\right) \eta_{\mathrm{m}}^{\mathrm{AB}}\left(V_{\mathrm{A}}^2-1\right)} \\
& C_{A_2 A_1}= \sqrt{T_{\mathrm{A}} \eta_{\mathrm{m}}^{\mathrm{AA}}\left(V_{\mathrm{A}}^2-1\right)}
\end{aligned}\right.,
\end{equation}
where $T_{\min} = \left(t_{\min}\right)^2$. The calculations of $t_{\min}$ and $\sigma_{\max}^2$ can be found in Equations~\ref{tmin} and~\ref{sigmamax}, respectively. By replacing the maximum likelihood estimators with their expectation values, $\hat{t} = \sqrt{T}$ and $\hat{\sigma}^2 = 1 + T \varepsilon$.

\subsection{Method for Calculating the Secret Key Rate}
\label{3.2}
After obtaining the covariance matrices $\gamma_{B_2 B_1 A_2 A_1}$ and $\left(\gamma_{B_2 B_1 A_2 A_1}\right)_{\mathrm{FAN}}$, we can calculate $I_{\mathrm{AB}}$ and $\left(S_{\mathrm{BE}}\right)_{\epsilon_{\mathrm{PE}}}$, and then evaluate the secret key rate using Eq.~\ref{keyrate}.
For the coherent-state heterodyne protocol with reverse reconciliation,

\begin{equation}
I_{\mathrm{AB}}=\log _2 \frac{V_{\mathrm{A}^{\mathrm{M}}}}{V_{\mathrm{A}^{\mathrm{M}} \mid \mathrm{B}}},
\end{equation}
where $V_{\mathrm{A}^{\mathrm{M}}}$ and $V_{\mathrm{A}^{\mathrm{M}} \mid \mathrm{B}}$ denote the variance of Alice's measurement outcomes and the conditional variance given Bob's measurement outcomes, respectively. From the covariance matrix shared by the two parties, we can obtain

\begin{equation}
\label{keypart1}
I_{\mathrm{AB}}=\log _2\left[\frac{T^2 T_{\mathrm{A}}(\chi+1)+T \chi+1+T\left(1-T_{\mathrm{A}}\right)\left(\eta_{\mathrm{m}}^{\mathrm{AB}} V_{\mathrm{A}}+1-\eta_{\mathrm{m}}^{\mathrm{AB}}\right)}{T^2 T_{\mathrm{A}}(\chi+1)+T \chi+1+T\left(1-T_{\mathrm{A}}\right)}\right].
\end{equation}

The information leaked from Bob to Eve, $\left(S_{\mathrm{BE}}\right)_{\epsilon_{\mathrm{PE}}}$, can be evaluated using the Holevo bound,

\begin{equation}
\left(S_{\mathrm{BE}}\right)_{\epsilon_{\mathrm{PE}}}=[S(\mathrm{E})]_{\epsilon_{\mathrm{PE}}}-\left[S\left(\mathrm{E} \mid x_{\mathrm{B}}, p_{\mathrm{B}}\right)\right]_{\epsilon_{\mathrm{PE}}},
\end{equation}
where $[S(\mathrm{E})]_{\epsilon_{\mathrm{PE}}}$ denotes Eve's von Neumann entropy, and $\left[S\left(\mathrm{E} \mid x_{\mathrm{B}}, p_{\mathrm{B}}\right)\right]_{\epsilon_{\mathrm{PE}}}$ denotes Eve's conditional von Neumann entropy given Bob's measurement outcomes.

After Eve purifies the entire system, we have $[S(\mathrm{E})]_{\epsilon_{\mathrm{PE}}}=[S(\mathrm{AB})]_{\epsilon_{\mathrm{PE}}}$. The entropy of the Gaussian state $\mathrm{AB}$ can then be calculated from its corresponding covariance matrix $\left(\gamma_{B_2 B_1 A_2 A_1}\right)_{\mathrm{FAN}}$,
\begin{equation}
\label{keypart2}
[S(\mathrm{E})]_{\epsilon_{\mathrm{PE}}}=\sum_{i=1}^7 G\left(\lambda_i\right),
\end{equation}

\begin{equation}
G\left(\lambda_i\right)=\frac{\lambda_i+1}{2} \log \frac{\lambda_i+1}{2}-\frac{\lambda_i-1}{2} \log \frac{\lambda_i-1}{2},
\end{equation}
where $\lambda_i$ denotes the symplectic eigenvalues of the matrix $\left(\gamma_{B_2 B_1 A_2 A_1}\right)_{\mathrm{FAN}}$.

The evaluation of Eve's conditional entropy given Bob's measurement outcomes, $\left[S\left(\mathrm{E} \mid x_{\mathrm{B}}, p_{\mathrm{B}}\right)\right]_{\epsilon_{\mathrm{PE}}}$, requires Bob's measurement results.
When Bob performs the data processing $x_{\mathrm{B}}=x_{\mathrm{B}_{2 \xi_{\mathrm{B}}}}-k x_{\mathrm{B}_{1 \xi_{\mathrm{B}}}}$ and $p_{\mathrm{B}}=p_{\mathrm{B}_{2 \xi_{\mathrm{B}}}}+k p_{\mathrm{B}_{1 \xi_{\mathrm{B}}}}$, the corresponding operations can be equivalently described by applying a symplectic transformation $\gamma_k$ to obtain a new covariance matrix,

\begin{equation}
\left(\gamma_{B_4 B_3 B_5 B_6 A_2 A_1}\right)_{\mathrm{FAN}}=\left[\gamma_k \oplus \gamma_k \oplus \mathrm{I}_2\right]\left(\gamma_{B_{2 X} B_{1 X} B_{1 p} B_{2 P} A_2 A_1}\right)_{\mathrm{FAN}}\left[\gamma_k \oplus \gamma_k \oplus \mathrm{I}_2\right]^{\mathrm{T}},
\end{equation}

\begin{equation}
\gamma_k=\left(\begin{array}{cccc}
1 & 0 & -k & 0 \\
0 & 1 & 0 & 0 \\
0 & 0 & 1 & 0 \\
0 & k & 0 & 1
\end{array}\right),
\end{equation}

\begin{equation}
k=T \sqrt{T_{\mathrm{A}}} \sqrt{\eta_{\mathrm{m}}^{\mathrm{BB}}} \sqrt{\frac{V_{\mathrm{B}}-1}{V_{\mathrm{B}}+1}}.
\end{equation}

After Eve purifies the system, $\left[S\left(\mathrm{E} \mid x_{\mathrm{B}}, p_{\mathrm{B}}\right)\right]_{\epsilon_{\mathrm{PE}}}=\left[S\left(B_3 B_5 A_2 A_{1 X} A_{1 P} \mid x_{\mathrm{B}}, p_{\mathrm{B}}\right)\right]_{\epsilon_{\mathrm{PE}}}$.
By performing heterodyne detection on the matrix $\left(\gamma_{B_4 B_3 B_5 B_6 A_2 A_1}\right)_{\mathrm{FAN}}$,

\begin{equation}
\left(\gamma_{B_3 B_5 A_2 A_1}^{X_{\mathrm{B}}, P_{\mathrm{B}}}\right)_{\mathrm{FAN}}=\left[\gamma_{B_3 B_5 A_2 A_1}-C_{B 4}\left(X \gamma_{B 4} X\right)^{M P} C_{B 4}{ }^T-C_{B 6}\left(P \gamma_{B 6} P\right)^{M P} C_{B 6}{ }^T\right]_{\mathrm{FAN}},
\end{equation}
we can then calculate

\begin{equation}
\label{keypart3}
\left[S\left(\mathrm{E} \mid x_{\mathrm{B}}, p_{\mathrm{B}}\right)\right]_{\epsilon_{\mathrm{PE}}}=\sum_{i=1}^5 G\left(\lambda_i^{\prime}\right),
\end{equation}
where $\lambda_i^{\prime}$ denotes the symplectic eigenvalues of the matrix $\left(\gamma_{B_3 B_5 A_2 A_1}^{X_{\mathrm{B}}, P_{\mathrm{B}}}\right)_{\mathrm{FAN}}$.

By substituting Equations~\ref{keypart1}, \ref{keypart2}, and~\ref{keypart3} into Equation~\ref{keyrate}, the secret key rate formula can be obtained.

%%%%%%%%%%%%%%%%%%%%%%%%%%%%%%%%%%%%%%%%%%
\section{Simulation and Analysis}
\label{IV}
In this section, we present the simulation results of the improved two-way protocol and compare them with the one-way coherent-state heterodyne protocol.

\begin{figure}[H]
    \centering
    \includegraphics[width=0.65\linewidth]{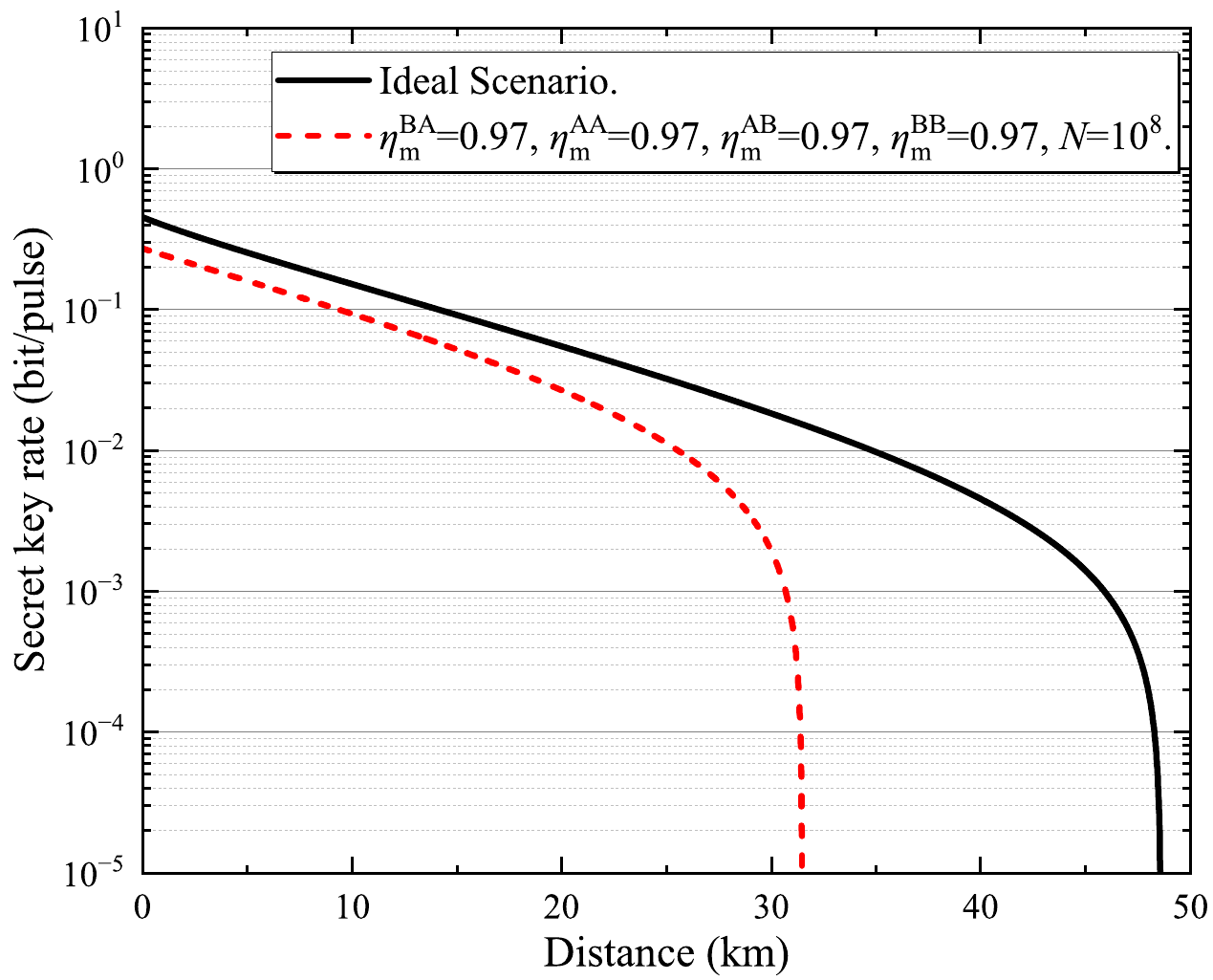}
    \caption{Comparison of the protocol performance between the ideal and practical scenarios. The black solid line represents the ideal scenario, while the red dashed line represents the practical scenario.}
    \label{FIG3}
\end{figure}

First, we compare the performance of the two-way protocol in the ideal scenario with the protocol in the practical scenario.
In the ideal case, the mode-matching coefficient is set to 1. Moreover, the analysis is carried out in the asymptotic regime.
In contrast, in the practical scenario, the four matching coefficients listed in Table~\ref{table1} must be taken into account simultaneously. To avoid the increased complexity arising from the variation of multiple matching coefficients, we assume that Alice and Bob employ the same types of lasers and detectors, and that the modulation format is pre-agreed to maintain temporal-mode consistency as much as possible. Therefore, for simplicity, the four matching coefficients are assumed to be identical in the simulations.
The simulation results are shown in Figure~\ref{FIG3}.
The channel transmittance is given by $T=10^{-\alpha L / 10}$, where a channel loss of $\alpha=0.2 ,\mathrm{dB}/\mathrm{km}$ is assumed and $L$ denotes the transmission distance. The variance of Bob's TCMSV state is $V_{\mathrm{B}}=20$, and that of Alice's TCMSV state is $V_{\mathrm{A}}=20$. The transmittance of the beam splitter used by Alice for interference is $T_{\mathrm{A}}=0.8$. The excess noise of the system is set to $\varepsilon=0.1$, and the reconciliation efficiency is $\beta_{\mathrm{R}}=0.95$.

In Figure~\ref{FIG3}, the black solid line represents the secret key rate of the improved two-way coherent-state heterodyne protocol as a function of transmission distance in the ideal scenario. The red dashed line represents the protocol performance when finite-size effects are taken into account and mode mismatch between the transmitter and the receiver is considered. It can be observed that the maximum transmission distance decreases from 48.27~km in the ideal scenario to 31.38~km, corresponding to a reduction of nearly 17~km. Therefore, when implementing the improved two-way protocol in practical scenario, the combined impact of finite-size effects and mode mismatch should be carefully considered.

\begin{figure}[H]
    \centering
    \includegraphics[width=0.65\linewidth]{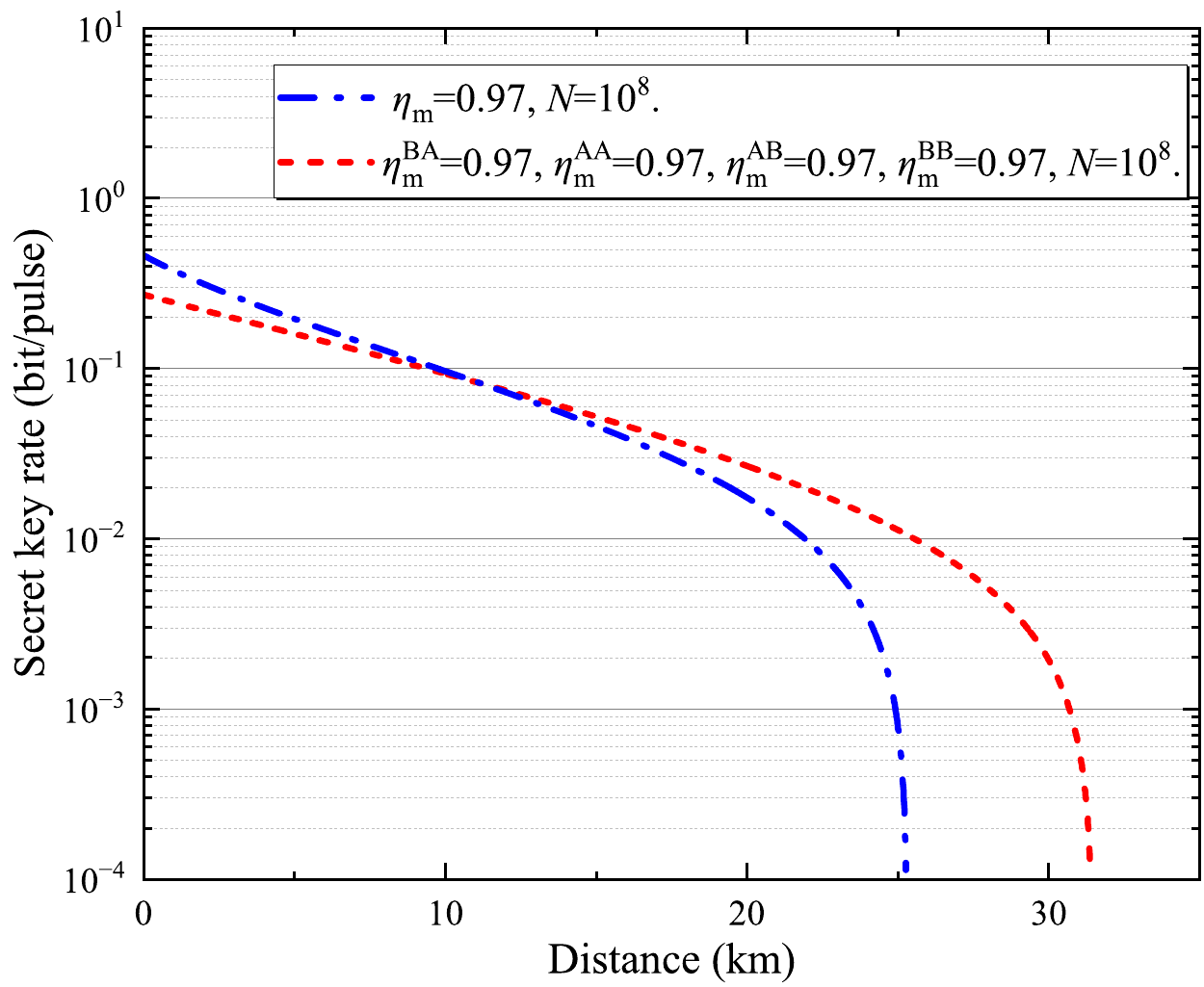}
    \caption{Comparison of the protocol performance between the improved two-way and one-way protocols. The red dashed line represents the improved two-way protocol, while the blue dash-dotted line represents the one-way protocol.}
    \label{FIG4}
\end{figure}

Although the performance of the improved two-way protocol degrades compared with that in the ideal scenario due to practical device nonidealities, it still outperforms the one-way protocol under the same parameter settings. To illustrate the performance advantage of the improved two-way protocol, we perform a comparative simulation analysis between the improved two-way and one-way protocols, and the results are shown in Figure~\ref{FIG4}.

In the continuous-mode scenario, finite-size effects are taken into account by setting the total number of exchanged signals to $N=10^8$. The mode-matching coefficients between the transmitter and the receiver are simultaneously set to 0.97. In addition, coherent states are prepared at the transmitter and heterodyne detection is performed at the receiver. Under these conditions, the red dashed line in Figure~\ref{FIG4} represents the improved two-way protocol, while the blue dash-dotted line represents the one-way protocol. It can be observed that the maximum transmission distance of the improved two-way protocol reaches 31.38~km, which is approximately 24\% higher than the one-way protocol, whose maximum transmission distance is 25.27~km. Moreover, by further reducing the excess noise or improving the mode matching between the transmitter and the receiver through digital signal processing techniques, the performance advantage of the improved two-way protocol over the one-way protocol can be further enhanced.

\begin{figure}[H]
    \centering
    \includegraphics[width=0.65\linewidth]{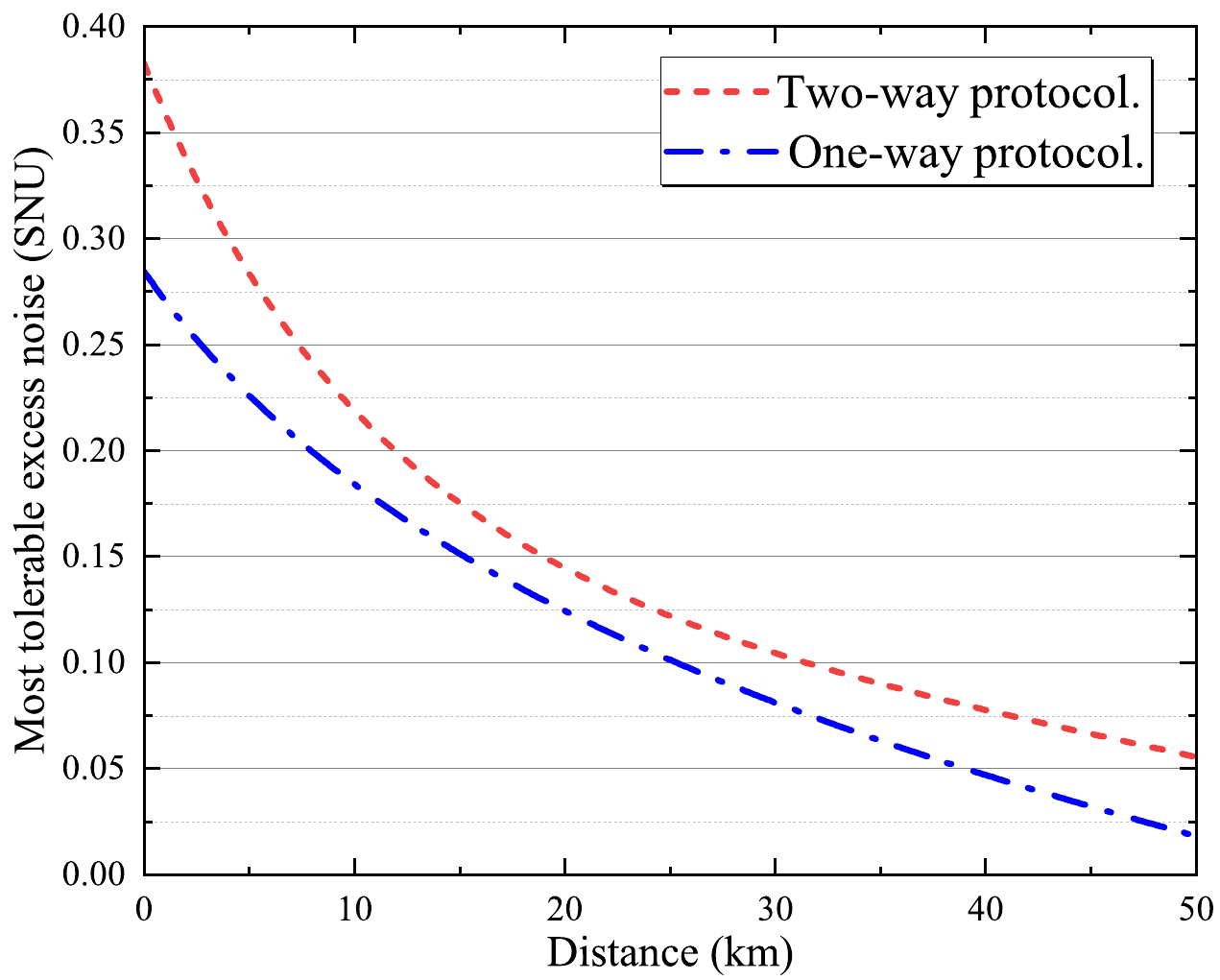}
    \caption{Comparison of the most tolerable excess noise between the improved two-way and one-way protocols. The red dashed line represents the improved two-way protocol, while the blue dash-dotted line represents the one-way protocol.}
    \label{FIG5}
\end{figure}

To further illustrate the advantage of the improved two-way protocol in tolerating excess noise, we compare the most tolerable excess noise of the improved two-way and one-way protocols. The simulation results are shown in Figure~\ref{FIG5}, where the red dashed line represents the improved two-way protocol and the blue dash-dotted line represents the one-way protocol.

The simulation results show that the improved two-way protocol exhibits a higher tolerance to excess noise than the one-way protocol over the entire transmission distance range. When the transmission distance is set to 30~km, the improved two-way protocol can tolerate an excess noise exceeding 0.1, whereas the maximum tolerable excess noise for the one-way protocol is only 0.08. As also observed in the simulation results shown in Figure~\ref{FIG4}, the one-way protocol is no longer feasible under these conditions. At a metropolitan-scale distance of 50~km, the maximum tolerable excess noise of the improved two-way protocol exceeds 0.05, while that of the one-way protocol is only 0.0182. In this case, the improved two-way protocol can tolerate approximately three times more excess noise than the one-way protocol. Moreover, as the transmission distance increases, system performance becomes increasingly sensitive to excess noise, making the advantages of the two-way protocol more pronounced.

%%%%%%%%%%%%%%%%%%%%%%%%%%%%%%%%%%%%%%%%%%
\section{Conclusions}
\label{V}
In this work, we characterize the continuous-mode optical fields in the improved two-way protocol by introducing temporal modes and establish a security analysis framework for the continuous-mode scenario based on adaptive normalization with calibrated shot-noise unit. 
In addition, finite-size effects on statistical fluctuations in parameter estimation were analyzed based on the central limit theorem and the maximum likelihood estimation method, leading to a tighter secret key rate for the improved two-way protocol.

Our numerical simulations indicate that, in practical scenarios, improving the mode matching between the transmitter and the receiver is essential for enhancing system performance. Under the same parameter settings, the improved two-way protocol consistently outperforms its one-way counterpart. As the transmission distance increases, the impact of excess noise on system performance becomes more pronounced, further highlighting the advantage of the improved two-way protocol in long-distance transmission.

This work provides a more realistic security and performance analysis for the improved two-way protocol, offering useful guidance for its practical implementation and performance optimization. In future digitalized systems, further performance improvements can be achieved by enhancing the mode-matching coefficient through optimized modulation formats and the incorporation of appropriate digital signal processing algorithms.

%%%%%%%%%%%%%%%%%%%%%%%%%%%%%%%%%%%%%%%%%%

%%%%%%%%%%%%%%%%%%%%%%%%%%%%%%%%%%%%%%%%%%
\vspace{6pt} 

%%%%%%%%%%%%%%%%%%%%%%%%%%%%%%%%%%%%%%%%%%
%% optional
%\supplementary{The following supporting information can be downloaded at:  \linksupplementary{s1}, Figure S1: title; Table S1: title; Video S1: title.}


% Only used for preprtints:
% \supplementary{The following supporting information can be downloaded at the website of this paper posted on \href{https://www.preprints.org/}{Preprints.org}.}

%\begin{tabularx}{\textwidth}{lll}
%\toprule
%\textbf{Name} & \textbf{Type} & \textbf{Description} \\
%\midrule
%S1 & Python script (.py) & Script of python source code used in XX \\
%S2 & Text (.txt) & Script of modelling code used to make Figure X \\
%S3 & Text (.txt) & Raw data from experiment X \\
%S4 & Video (.mp4) & Video demonstrating the hardware in use \\
%\bottomrule
%\end{tabularx}
%}

%%%%%%%%%%%%%%%%%%%%%%%%%%%%%%%%%%%%%%%%%%
\authorcontributions{
Conceptualization, Y.S., J.M., X.W. and Z.C.; 
Methodology, Y.S.; 
Software, Y.S.; 
Formal analysis, Y.S., X.W. and Z.C.; 
Investigation, Y.S.; 
Resources,  X.W., Z.C., S.Y. and H.G.; 
Writing---original draft preparation, Y.S.; 
Writing---review and editing, X.W., Z.C. and S.Y.; 
Visualization, Y.S. and J.M.; 
Supervision, X.W., Z.C., S.Y. and H.G.; 
Project administration, X.W. and Z.C.; 
Funding acquisition, X.W., Z.C., S.Y. and H.G. 
All authors have read and agreed to the published version of the manuscript.
}

\funding{
This work was funded by the National Natural Science Foundation of China, 
grant numbers 62371060, 62201012, 62001041 and 62571006, 
and by the Fund of State Key Laboratory of Information Photonics and Optical Communications, 
grant number IPOC2022ZT09.
}

\institutionalreview{Not applicable.}

\informedconsent{Not applicable.}

\dataavailability{Data sharing not applicable to this article as no datasets were generated or analyzed during the current study.}

\acknowledgments{Not applicable.}

\conflictsofinterest{The authors declare no conflicts of interest.} 

%%%%%%%%%%%%%%%%%%%%%%%%%%%%%%%%%%%%%%%%%%
%% Optional

%% Only for journal Encyclopedia
%\entrylink{The Link to this entry published on the encyclopedia platform.}

%\abbreviations{Abbreviations}{ ... }

%%%%%%%%%%%%%%%%%%%%%%%%%%%%%%%%%%%%%%%%%%
%% Optional
%\appendixtitles{no} % Leave argument "no" if all appendix headings stay EMPTY (then no dot is printed after "Appendix A"). If the appendix sections contain a heading then change the argument to "yes".
%\appendixstart
%\appendix
%\section[]{}

%%%%%%%%%%%%%%%%%%%%%%%%%%%%%%%%%%%%%%%%%%
%\isPreprints{}{% This command is only used for ``preprints''.
\begin{adjustwidth}{-\extralength}{0cm}

\reftitle{References}


% \bibliography{your_external_BibTeX_file}

%=====================================
% References, variant B: internal bibliography
%=====================================

% ACS format
\isAPAandChicago{}{%

}

% If authors have biography, please use the format below
%\section*{Short Biography of Authors}
%\bio
%{\raisebox{-0.35cm}{\includegraphics[width=3.5cm,height=5.3cm,clip,keepaspectratio]{Definitions/author1.pdf}}}
%{\textbf{Firstname Lastname} Biography of first author}
%
%\bio
%{\raisebox{-0.35cm}{\includegraphics[width=3.5cm,height=5.3cm,clip,keepaspectratio]{Definitions/author2.jpg}}}
%{\textbf{Firstname Lastname} Biography of second author}

% For the MDPI journals use author-date citation, please follow the formatting guidelines on http://www.mdpi.com/authors/references
% To cite two works by the same author: \citeauthor{ref-journal-1a} (\citeyear{ref-journal-1a}, \citeyear{ref-journal-1b}). This produces: Whittaker (1967, 1975)
% To cite two works by the same author with specific pages: \citeauthor{ref-journal-3a} (\citeyear{ref-journal-3a}, p. 328; \citeyear{ref-journal-3b}, p.475). This produces: Wong (1999, p. 328; 2000, p. 475)

%%%%%%%%%%%%%%%%%%%%%%%%%%%%%%%%%%%%%%%%%%
%% for journal Sci
%\reviewreports{\\
%Reviewer 1 comments and authors’ response\\
%Reviewer 2 comments and authors’ response\\
%Reviewer 3 comments and authors’ response
%}
%%%%%%%%%%%%%%%%%%%%%%%%%%%%%%%%%%%%%%%%%%
\PublishersNote{}
%\isPreprints{}{% This command is only used for ``preprints''.
\end{adjustwidth}
\end{document}